\documentstyle[twocolumn,prl,aps,epsf]{revtex} 
\newcommand{\be}{\begin{eqnarray}}
\newcommand{\ee}{\end{eqnarray}}

\begin{document}
\draft
\title{Dispersion relation formalism for virtual Compton 
scattering and the generalized polarizabilities of the nucleon}

\author{B. Pasquini$^1$, D. Drechsel$^2$, M. Gorchtein$^2$, 
A. Metz$^3$, and M. Vanderhaeghen$^2$}
\address{$^1$ ECT*, European Center for Theoretical Studies 
in Nuclear Physics and Related Areas, Trento, 
{\rm and} INFN, Trento, Italy}
\address{$^2$ Institut f\"ur Kernphysik, Johannes Gutenberg-Universit\"at,
  D-55099 Mainz, Germany}
\address{$^3$ CEA-Saclay, DAPNIA/SPhN, F-91191 Gif-sur-Yvette, France}
\date{\today}
\maketitle

\begin{abstract}
A dispersion relation formalism for the virtual Compton scattering
(VCS) reaction on the proton is presented,
which for the first time allows a dispersive evaluation of 4 generalized
polarizabilities at a four-momentum transfer $Q^2 \leq$ 0.5 GeV$^2$. 
The dispersive integrals are calculated using a 
state-of-the-art pion photo- and electroproduction analysis. 
The dispersion formalism provides a new tool to analyze VCS
experiments above pion threshold, thus increasing the sensitivity to the
generalized polarizabilities of the nucleon. 
\end{abstract}
\narrowtext

Over the past years, the virtual Compton scattering (VCS) 
process on the proton, accessed through the $e p \to e p \gamma$
reaction, has become 
a powerful and precise tool to provide new information on the internal
structure of the nucleon \cite{GuiVdh}. VCS has been
shown to be of particular interest not only at low outgoing 
photon energies where one probes nucleon-core excitations 
and pion-cloud contributions to so-called generalized
polarizabilities, but also at high energy and momentum transfers, 
where one is sensitive to a new type of parton distributions, 
generalizing the information obtained 
from inclusive deep-inelastic scattering. 
\newline
\indent
In the low energy regime below pion threshold, the outgoing photon in
the VCS process plays the role of a quasi-constant applied 
electromagnetic dipole field and, through electron scattering, 
one measures the spatial distribution of the nucleon response 
to this applied field \cite{GuiVdh}. The response is
parametrized in terms of 6 generalized polarizabilities (GP's)
\cite{Gui95,Dre97}, which are functions of the  
square of the virtual photon four-momentum $Q^2$. The GP's provide valuable
non-perturbative nucleon structure information, and have been
calculated in different approaches 
\cite{Vdh96,metz96,Liu96,Hem97,Pas98,Pas00}. 
In particular, the GP's teach us about the interplay
between nucleon-core excitations and pion-cloud effects. 
\newline
\indent
The first dedicated VCS experiment has been
performed at MAMI \cite{Roc00} and two combinations 
of GP's have been determined at $Q^2$ = 0.33 GeV$^2$. 
Further VCS experiments are underway at lower $Q^2$ 
at MIT-Bates \cite{Sha97} and at higher $Q^2$ at JLab \cite{Bert93}.  
\newline
\indent
At present, VCS experiments at low outgoing photon energies 
are analyzed in terms of a low-energy expansion as proposed in
\cite{Gui95}, assuming that the non-Born response of the system 
to the quasi-constant electromagnetic field of the low energetic 
photon is proportional to the GP's. 
As the sensitivity of the VCS cross sections to the GP's 
grows with the photon energy, it is
advantageous to go to higher photon energies, provided one can keep the
theoretical uncertainties under control when crossing the pion
threshold. The situation can be compared to real Compton
scattering (RCS), for which one uses a dispersion relation formalism
\cite{lvov97,Dre99} to extract the polarizabilities at energies 
above pion threshold, with generally larger effects on the observables.  
The aim of the present work is to provide such 
a dispersion formalism for VCS, as a tool to analyze 
VCS experiments at higher energies in order to extract 
the GP's from data over a larger energy range. 
It will be shown that the same formalism also provides for the first
time a dispersive evaluation of 4 GP's.
\newline
\indent
To calculate the VCS process, we start from the helicity amplitudes~:
\begin{equation}
T_{\lambda' s'; \, \lambda s} = -e^2 \varepsilon_\mu(q, \lambda) \, 
\varepsilon^{'*}_\nu(q', \lambda') \,
\bar u(p', s')  {\mathcal M}^{\mu \nu} u(p, s),
\label{eq:matrixele}
\end{equation}
with $e$ the electric charge, $q$ ($q'$) 
the four-vectors of the virtual (real) photon in the VCS process, 
and $p$ ($p'$) the four-momenta of the initial (final)
nucleons respectively. The nucleon helicities are
denoted by $s, s' = \pm 1/2$, and $u, \bar u$ are the nucleon spinors. 
The initial virtual photon has helicity $\lambda = 0, \pm 1$ and 
polarization vector $\varepsilon_\mu$, whereas 
the final real photon has helicity $\lambda' = \pm 1$ and 
polarization vector $\varepsilon^{'}_\nu$. The VCS process is
characterized by 12 independent helicity amplitudes 
$T_{\lambda' s'; \, \lambda s}$.   
\newline
\indent
The VCS tensor ${\mathcal M}^{\mu \nu}$ in
Eq.~(\ref{eq:matrixele}) is then expanded into a basis of 
12 independent gauge invariant tensors $\rho^{\mu \nu}_i$,  
\begin{equation}
{\mathcal M}^{\mu \nu} \;=\; \sum_{i = 1}^{12} 
\; F_i(Q^2, \nu, t) \, \rho^{\mu \nu}_i \;, 
\label{eq:nonborn}
\end{equation}
as introduced in \cite{Dre97} (starting from the amplitudes of \cite{Tar75}). 
The amplitudes $F_i$ in Eq.~(\ref{eq:nonborn}) contain all nucleon
structure information and are functions of 3 invariants for the
VCS process~: $Q^2 \equiv - q^2$, $\nu = (s - u)/(4 M_N)$  which is odd under 
$s \leftrightarrow u$ crossing, and $t$. 
The Mandelstam invariants $s$, $t$ and $u$ for VCS are defined by 
$s = (q + p)^2$, $t = (q - q')^2$, and $u = (q - p')^2$, with the
constraint $s + t + u = 2 M_N^2 - Q^2$, and $M_N$ is the nucleon mass.
\newline
\indent
Nucleon crossing combined with charge conjugation provides the
following constraints on the amplitudes $F_i$ 
\footnote{We have redefined 4 of the 12 invariant amplitudes
  of \cite{Dre97} by dividing them through $\nu$, such that
  all of them are even functions of $\nu$. This simplifies the
  formalism since only one type of dispersion integrals 
  needs to be considered then.} 
at arbitrary virtuality $Q^2$ :
\begin{equation}
F_i \left( Q^2, -\nu, t \right) 
= F_i\left( Q^2, \nu, t \right) \hspace{.5cm}
(i = 1,...,12).
\label{eq:crossing}  
\end{equation}
In a next step, the VCS tensor ${\mathcal M}^{\mu \nu}$ 
at low outgoing photon
energies is separated into Born ($B$) and non-Born ($NB$) parts, 
as described in \cite{Gui95}. 
In the Born process, the virtual photon is
absorbed on a nucleon and the intermediate state remains a nucleon,
whereas the non-Born process contains all nucleon excitations 
and meson-loop contributions, and is parametrized
through 6 GP's.  
With the choice of the tensor basis of \cite{Dre97}, 
the resulting non-Born amplitudes $F^{NB}_i$ ($i$ = 1,...,12)
are free of all kinematical singularities and constraints. 
\newline
\indent
Assuming further analyticity and
an appropriate high-energy behavior, the non-Born amplitudes 
$F^{NB}_i(Q^2, \nu, t)$ fulfill 
unsubtracted dispersion relations (DR's) 
with respect to the variable $\nu$ at
fixed $t$ and fixed virtuality $Q^2$~:
\begin{equation}
{\mathrm Re} F_i^{NB}(Q^2, \nu, t) = 
{2 \over \pi}  {\mathcal P} \int_{\nu_{thr}}^{+ \infty} d\nu'  
{{\nu' \, {\mathrm Im}_s F_i(Q^2, \nu',t)} \over {\nu'^2 - \nu^2}},
\label{eq:unsub} 
\end{equation}
with ${\mathrm Im}_s F_i$ the discontinuities 
across the $s$-channel cuts of the VCS process. 
Since pion production is the first inelastic channel, 
$\nu_{thr} = m_\pi + (m_\pi^2 + t/2 + Q^2/2)/(2 M_N)$, 
where $m_\pi$ denotes the pion mass. 
\newline
\indent
The unsubtracted DR's of Eq.~(\ref{eq:unsub}) require 
that at sufficiently high energies ($\nu \rightarrow \infty$ 
at fixed $t$ and fixed $Q^2$) the
amplitudes ${\mathrm Im}_s F_i(Q^2,\nu,t)$ ($i$ = 1,...,12) 
drop fast enough such that the
integrals are convergent and the contributions from
the semi-circle at infinity can be neglected. 
The high-energy behavior of the amplitudes $F_i$ is deduced 
from the high-energy behavior of the VCS helicity amplitudes 
of Eq.~(\ref{eq:matrixele}). 
After some algebra, we obtain the following Regge limit for 
$\nu \to \infty$, at fixed $t$ and $Q^2$~: 
\begin{eqnarray}
F_{1}, \; F_{5} \;&\sim&\; Q^2 \, \nu^{\alpha_P(t) - 1} \;,\hspace{.35cm}
\nu^{\alpha_M(t)} \; , 
\label{eq:reggevcs1}\\
F_{7}  
\;&\sim&\; \nu^{\alpha_P(t) - 2} \;,\hspace{.35cm} \nu^{\alpha_M(t) - 1} \; , 
\label{eq:reggevcs2}\\
F_{2}, F_{3}, \left(F_{5} + 4 F_{11}\right), \nonumber\\
F_{6}, F_{8}, F_{9}, F_{10}, F_{12} 
\;&\sim&\; \nu^{\alpha_P(t) - 2} \;,\hspace{.35cm} \nu^{\alpha_M(t) - 2} \; , 
\label{eq:reggevcs3} \\
F_{4}  
\;&\sim&\; \nu^{\alpha_P(t) - 4} \;,\hspace{.35cm} \nu^{\alpha_M(t) - 3} \; . 
\label{eq:reggevcs4}
\end{eqnarray}
In Eqs.~(\ref{eq:reggevcs1})-(\ref{eq:reggevcs4}), 
we have separately indicated the high energy
behavior originating from the ``pomeron'' 
\footnote{The pomeron contributes to the sum of 
helicity conserving amplitudes $T_{1 \,{1 \over 2}; 1 \,{1 \over 2}} +
T_{-1 \,{1 \over 2}; -1 \,{1 \over 2}}$, i.e. gives the same 
amplitude for parallel or antiparallel orientation of the helicities.}
(with Regge trajectory $\alpha_P(t)$, and $\alpha_P(0) \approx $ 1.08) 
and from $t$-channel meson-exchange contributions 
(with Regge trajectory $\alpha_M(0) \lesssim 0.5$ ). 
It then follows from Eqs.~(\ref{eq:reggevcs1})-(\ref{eq:reggevcs4}) 
for two amplitudes, $F_1$ and $F_5$, 
that an unsubtracted dispersion integral as in Eq.~(\ref{eq:unsub})
does not exist, whereas the other 10 amplitudes 
on the {\it lhs} of Eqs.~(\ref{eq:reggevcs2})-(\ref{eq:reggevcs4}) can
be evaluated through unsubtracted dispersion integrals. 
This situation is similar as for RCS, 
where 2 of the 6 invariant amplitudes cannot be evaluated 
by unsubtracted dispersion relations either \cite{lvov97}.
\newline
\indent 
To construct the VCS amplitudes $F_1$ and $F_5$ in an unsubtracted 
dispersion framework, one could proceed in an analogous way as 
has been proposed by L'vov \cite{lvov97} in the case of RCS.  
The unsubtracted dispersion integrals for $F_1$ and $F_5$ 
are evaluated along the real $\nu$-axis in a finite range 
$-\nu_{max}\leq\nu\leq+\nu_{max}$ (with $\nu_{max}\approx$ 1.5~GeV).   
The integral along a semi-circle of finite radius $\nu_{max}$ 
in the complex $\nu$-plane is described by  
the asymptotic contribution $F_i^{as}$, 
which is parametrized by $t$-channel poles (e.g. for $Q^2$ = 0,
$F_1^{as}$ corresponds to $\sigma$-exchange, and $F_5^{as}$ corresponds
to $\pi^0$-exchange). 
Since the parametrization of the asymptotic parts amounts to some 
phenomenology, 
we will limit ourselves in the present work to those 10 amplitudes 
$F_i$ ($i \neq$ 1, 5) for which the integrals 
in Eq.~(\ref{eq:unsub}) converge, 
and which do not involve such asymptotic contributions.  
A full study of VCS observables within a dispersion
formalism - requiring, of course, all 12 amplitudes $F_i$, and
a parametrization of the two asymptotic contributions -
will be considered in a future work. 
\newline
\indent 
We next consider the non-Born VCS tensor 
at low energy ($|\vec q^{\; '}| \to 0$) 
but at arbitrary three-momentum $| \vec q \,|$ of the virtual photon.  
In this limit, it has been shown \cite{Gui95,Dre97} that 
the non-Born term can be parametrized by 6 generalized
polarizabilities (GP's), 
which are functions of $| \vec q \,|$ and which are 
denoted by $P^{(\rho' \, L', \rho \,L)S}(| \vec q \,|)$. 
In this notation, $\rho$ ($\rho'$) refers to the
electric (2), magnetic (1) or longitudinal (0) nature of the initial 
(final) photon, $L$ ($L' = 1$) represents the angular momentum of the
initial (final) photon, and $S$ differentiates between the 
spin-flip ($S=1$) and non spin-flip ($S=0$) 
character of the transition at the nucleon side.   
A convenient choice for the 6 GP's has been proposed in \cite{GuiVdh}, 
\begin{eqnarray}
&&P^{(01,01)0}(| \vec q \,|),\;\;\; P^{(11,11)0}(| \vec q \,|), \;
\label{eq:defgpunpol} \\
&&P^{(01,01)1}(| \vec q \,|),\;\;\; P^{(11,11)1}(| \vec q \,|),\; \\
&&P^{(01,12)1}(| \vec q \,|),\;\;\; P^{(11,02)1}(| \vec q \,|),  
\label{eq:defgppol}
\end{eqnarray}
which reduces to the following expressions in the real photon 
limit ($| \vec q \,| = 0$) \cite{Dre97}~: 
\begin{eqnarray}
&&P^{(01,01)0}(0) \sim \alpha,\;\;\; 
P^{(11,11)0}(0) \sim \beta, \; \\
&&P^{(01,01)1}(0) = 0,\;\;\; P^{(11,11)1}(0) = 0,\; \\
&&P^{(01,12)1}(0) \sim \gamma_3,  \;\;\;
P^{(11,02)1}(0) \sim \gamma_2 + \gamma_4, 
\end{eqnarray}
where $\alpha$ ($\beta$) are the electric (magnetic) polarizabilities, 
and $\gamma_2, \gamma_3, \gamma_4$ are 3 of the 4 
spin polarizabilities of RCS. 
\newline
\indent
In terms of the VCS invariants, 
the limit $|\vec q^{\; '}| \to 0$ at finite $| \vec q \,|$ corresponds to 
$\nu \to 0$ and $t \to -Q^2$ at finite $Q^2$. 
One can therefore express the GP's in terms of
the VCS amplitudes $F_i$ at the point $\nu = 0$, $t = -Q^2$ at
finite $Q^2$, for which we introduce the shorthand~:
$\bar F_i(Q^2) \;\equiv\; F_i^{NB} \left(Q^2, \nu = 0, t = - Q^2 \right)$.
The relations between the GP's and the $\bar F_i(Q^2)$ can be found in
\cite{Dre97}. 
\newline
\indent 
From the high-energy behavior for the VCS invariant amplitudes, 
it follows that one can evaluate the $\bar F_i$ (for $i \neq$ 1, 5) 
through the unsubtracted DR's
\begin{equation}
\bar F_i(Q^2) \;=\; 
{2 \over \pi} \; \int_{\nu_{thr}}^{+ \infty} d\nu' \; 
{{{\mathrm Im}_s F_i(Q^2, \nu',t = - Q^2)} \over {\nu'}}\;.
\label{eq:sumrule} 
\end{equation} 
Unsubtracted DR's for the GP's will 
therefore hold for those combinations of GP's that do not  
depend upon the amplitudes $\bar F_1$ and $\bar F_5$. 
We note however that $\bar F_5$ can appear in the combination 
$\bar F_5 + 4 \, \bar F_{11}$, which has the high-energy behavior 
of Eq.~(\ref{eq:reggevcs3}) leading to a convergent integral. 
Among the 6 GP's, we find the following 4 combinations 
that do not depend upon $\bar F_1$ and $\bar F_5$~:
\begin{eqnarray}
&&P^{\left(0 1, 0 1\right)0} + {1 \over 2}  
P^{\left(1 1, 1 1\right)0} = 
{{-2} \over {\sqrt{3}}}\,\left( {{E + M_N} \over
    E}\right)^{1/2} M_N\,\tilde q_0\, \nonumber\\
&&\hspace{1.1cm} \times 
\left\{ {{| \vec q \,|^2} \over {\tilde q_0^2}}\, \bar F_2 + 
\left( 2 \, \bar F_6 + \bar F_9 \right) - \bar F_{12} \right\}, 
\label{eq:gpdisp1} \\
&&P^{\left(0 1, 0 1\right)1} = 
{1 \over {3 \sqrt{2}}}\,\left( {{E + M_N} \over E}\right)^{1/2} 
\,\tilde q_0\, \nonumber\\
&&\hspace{1.1cm}\times 
\left\{ \left( \bar F_5 + \bar F_7 + 4\, \bar F_{11} \right) 
+ 4 \, M_N \, \bar F_{12} \right\}, 
\label{eq:gpdisp2} \\
&&P^{\left(0 1, 1 2\right)1} - {1 \over {\sqrt{2} \, \tilde q_0}}  
P^{\left(1 1, 1 1\right)1} = 
{1 \over {3}} \left( {{E + M_N} \over E}\right)^{1/2} 
{{M_N \, \tilde q_0} \over {| \vec q \,|^2}} \nonumber\\
&&\hspace{1.1cm}\times
\left\{ \left( \bar F_5 + \bar F_7 + 4\, \bar F_{11} \right) 
+ 4 \, M_N \left( 2 \, \bar F_6 + \bar F_9 \right) \right\}, 
\label{eq:gpdisp3} \\
&&P^{\left(0 1, 1 2\right)1} +  
{{\sqrt{3}} \over {2}}  P^{\left(1 1, 0 2\right)1} =
{1 \over {6}} \left( {{E + M_N} \over E}\right)^{1/2} \, 
{{\tilde q_0} \over {| \vec q \,|^2}} \nonumber \\
&&\hspace{.8cm} \times 
\left\{ \tilde q_0 \left( \bar F_5 + \bar F_7 + 4\, \bar F_{11} \right) 
+ 8 \, M_N^2 \left( 2 \, \bar F_6 + \bar F_9 \right) \right\}, 
\label{eq:gpdisp4} 
\end{eqnarray}
where $E = \sqrt{|\vec q \,|^2 + M_N^2}$ denotes 
the initial proton c.m. energy, and  
$\tilde q_0 = M_N - E$ the virtual photon c.m. energy in the limit 
$|\vec q^{\; '}|$ = 0. Unfortunately, the 4 combinations of GP's of 
Eqs.~(\ref{eq:gpdisp1})-(\ref{eq:gpdisp4}) can at present not yet
be compared with the data. In particular, the only unpolarized experiment 
\cite{Roc00} measured two structure functions which 
cannot be evaluated in an unsubtracted DR formalism, 
as they contain in addition to 
$P^{\left(0 1, 0 1\right)0} + 1/2 P^{\left(1 1, 1 1\right)0}$ 
of Eq.~(\ref{eq:gpdisp1}), which
is proportional to $\alpha + \beta$ at $Q^2$ = 0, also the
generalization of $\alpha - \beta$. 
\newline
\indent
The 4 combinations of GP's on the {\it lhs} of 
Eqs.~(\ref{eq:gpdisp1})-(\ref{eq:gpdisp4}) can then be evaluated by 
unsubtracted DR's, from the dispersion integrals
of Eq.~(\ref{eq:sumrule}) for the $\bar F_i(Q^2)$. 
To this end, the imaginary parts ${\mathrm Im}_s F_i$ 
in Eq.~(\ref{eq:sumrule}) have to be calculated by use of unitarity. For the
VCS helicity amplitudes of Eq.~(\ref{eq:matrixele})
(denoted for short by $T_{fi}$), the unitarity equation reads~:
\begin{equation}
\label{s-unit}
2\,{\rm Im}_s\,T_{fi}=
\sum_X (2\pi)^4 \delta^4(P_X-P_i)T^{\dagger}_{X f }\,T_{X i} \;,
\end{equation}
where the sum runs over all possible intermediate states 
$X$ that can be formed. 
In our present calculation, we saturate the dispersion integrals of
Eq.~(\ref{eq:sumrule}) by the dominant contribution of the $\pi N$
intermediate states. For the pion photo- and electroproduction
helicity amplitudes in the range $Q^2 \leq$ 0.5 GeV$^2$, we  
use the phenomenological analysis of MAID \cite{maid00}, 
which contains both resonant and non-resonant pion production mechanisms.
\begin{figure}[h]
\epsfxsize=8.5 cm
\epsfysize=13.7 cm
\centerline{\hspace{.5cm}\epsffile{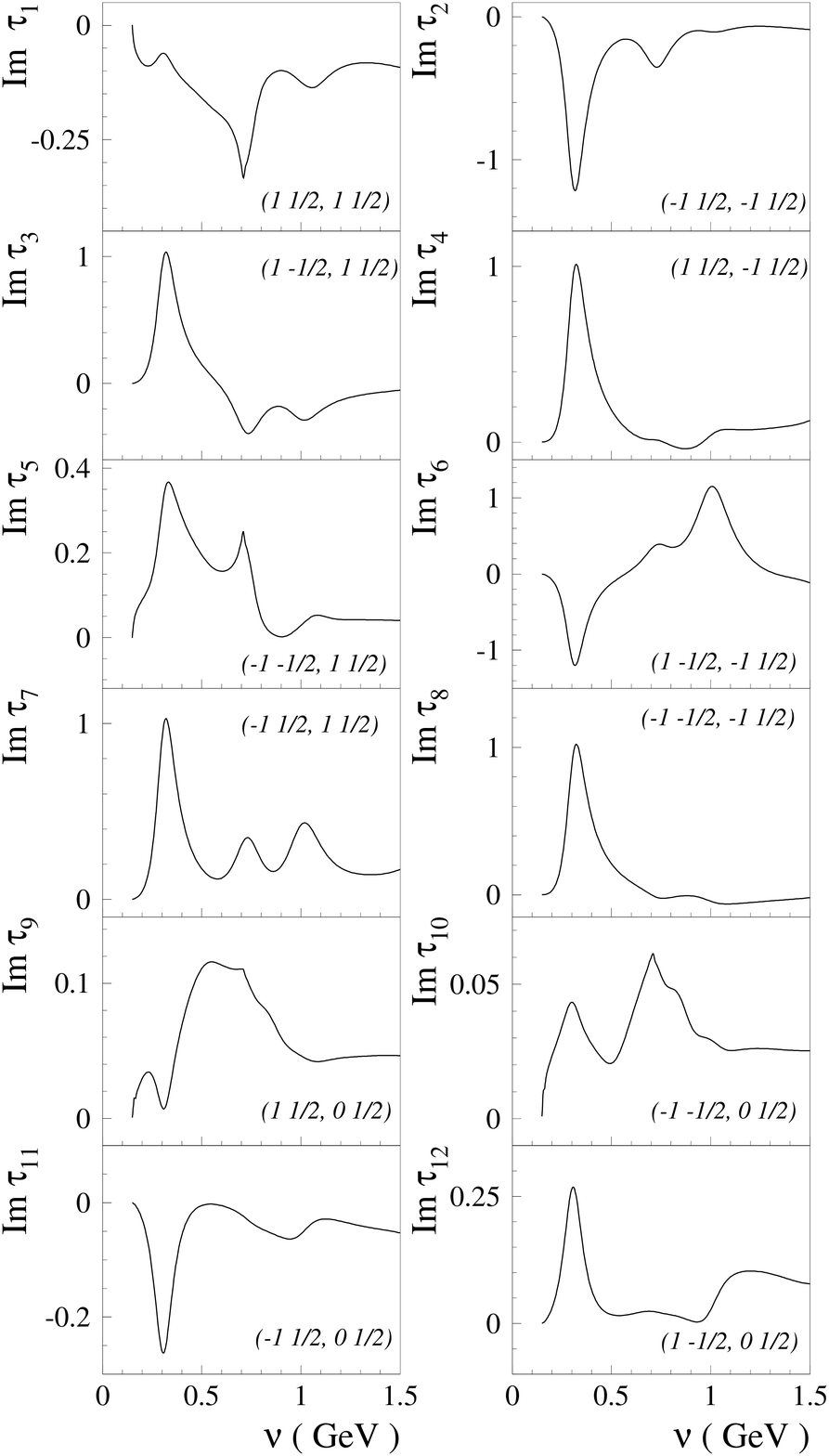}}
\caption[]{\small Imaginary parts of the VCS helicity amplitudes 
of the proton at $-t$ = $Q^2$ = 0.3 GeV$^2$, 
evaluated with $\pi^+ n$ and $\pi^0 p$ intermediate states. The labels 
$(\lambda' s', \lambda s)$ indicate the corresponding helicities. 
}
\label{fig:tau}
\end{figure}
In Fig.~\ref{fig:tau}, we show the imaginary parts
of the 12 VCS reduced helicity amplitudes $\tau_i$ 
as calculated with $\pi N$ intermediate states.  
These reduced amplitudes are obtained from
the full VCS helicity amplitudes of Eq.~(\ref{eq:matrixele})
by dividing out a common angular factor~:
\begin{equation}
T_{\lambda' s'; \lambda s}
=(\cos \theta/2)^{|\Lambda+\Lambda'|} \;
(\sin \theta/2)^{|\Lambda-\Lambda'|} \;
\tau_i \, .
\label{eq:redhel}
\end{equation}
In Eq.~(\ref{eq:redhel}), 
the total helicities in the initial and final states are denoted 
by $\Lambda=\lambda - s$ and $\Lambda'=\lambda' - s'$ respectively, 
and the correspondence between the 12 $\tau_i$ and the helicity labels
is given in Fig.~\ref{fig:tau}. 
The imaginary parts were
calculated in two different ways, first through the helicity amplitudes
as expressed in Eq.~(\ref{s-unit}). The sum over 
the final states denoted by $X$ contains a
phase-space integral for the $\pi N$ intermediate state, which is
then performed numerically. In a second calculation, the
helicity amplitudes are first decomposed into a multipole series,  
and the unitarity equation is then implemented for the $\pi N$ multipoles. 
It was found that the partial
wave expansion up to orbital angular momentum $ l \leq 3$, 
is already in very good agreement with the numerical integration, 
thus providing a valuable cross-check on the numerical calculation.  
\begin{figure}[h]
\epsfxsize=8.25 cm
\epsfysize=8.5 cm
\centerline{\epsffile{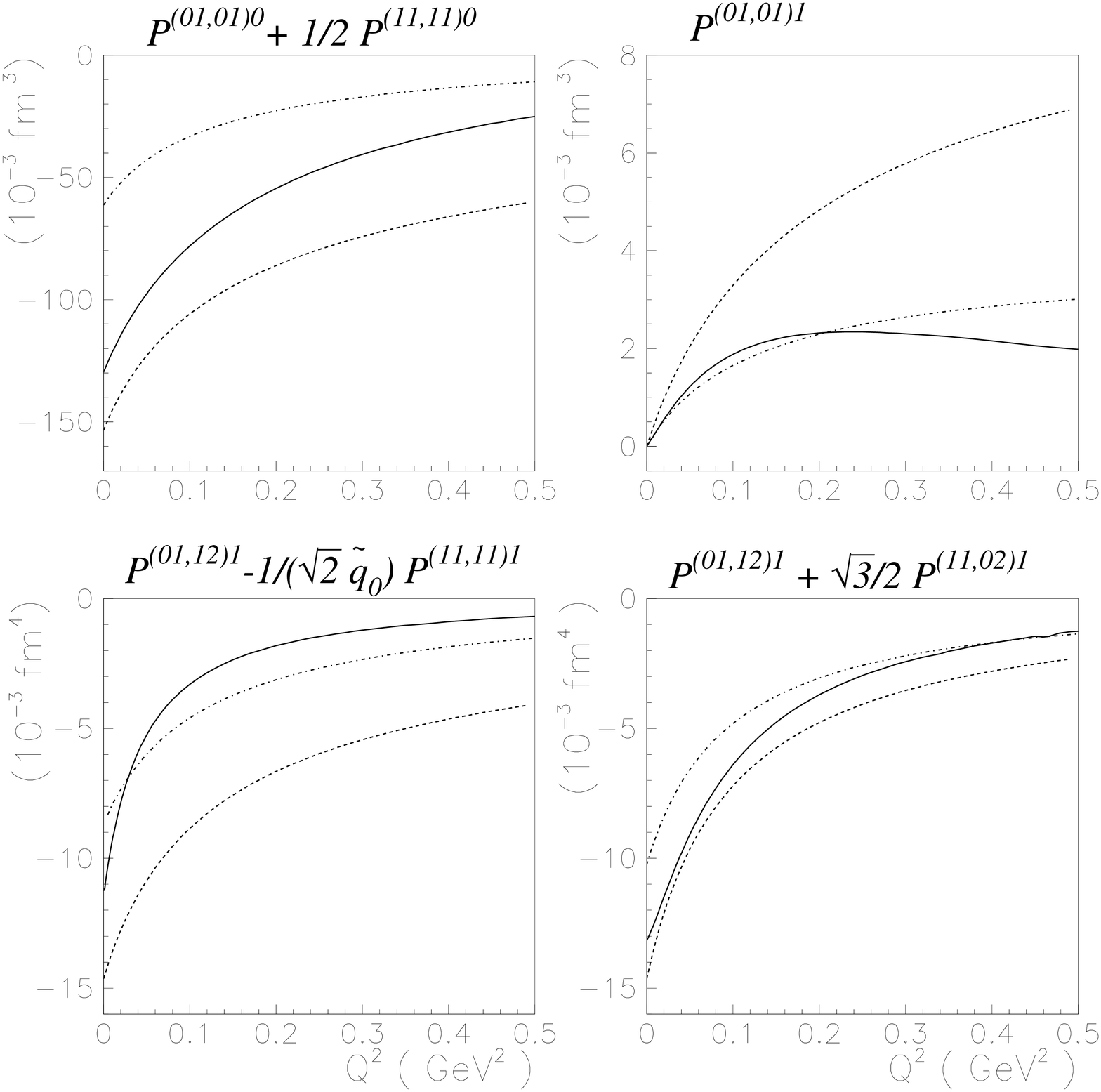}}
\caption[]{\small Dispersion results for 4 of the generalized 
polarizabilities of the proton 
(full curves), compared with results of $O(p^3)$ 
HBChPT \cite{Hem97} (dashed curves) and the linear $\sigma$-model
\cite{metz96} (dashed-dotted curves).}
\label{fig:polcomb}
\end{figure}
In Fig.~\ref{fig:polcomb}, we show the 
results for the 4 combinations of GP's 
of Eqs.~(\ref{eq:gpdisp1})-(\ref{eq:gpdisp4}) 
in the DR formalism, and compare them to the results of the 
$O(p^3)$ heavy-baryon chiral perturbation theory (HBChPT) \cite{Hem97}
and the linear $\sigma$-model \cite{metz96}. 
The $\pi N$ contribution to the sum 
$P^{\left(0 1, 0 1\right)0} + 1/2 P^{\left(1 1, 1 1\right)0}$ 
gives only about 80\% of
the Baldin sum rule \cite{Bab98}, because of a non-negligible
high-energy contribution (of heavier intermediate states) to the  
photoabsorption cross section entering the sum rule, 
which is not estimated here. For the 3 combinations
of Eqs.~(\ref{eq:gpdisp2})-(\ref{eq:gpdisp4}) of spin polarizabilities, we
expect the dispersive estimates with $\pi N$ states to
provide a rather reliable guidance. By comparing our results with those 
of HBChPT at $O(p^3)$, we note a rather good agreement for  
$P^{\left(0 1, 1 2\right)1} + \sqrt{3}/2 P^{\left(1 1, 0 2\right)1}$,  
whereas for $P^{\left(0 1, 0 1\right)1}$ and 
$P^{\left(0 1, 1 2\right)1} - 1/(\sqrt{2} \, \tilde q_0) 
P^{\left(1 1, 1 1\right)1}$,
the dispersive results drop much faster with $Q^2$. This trend
is also seen in the relativistic linear $\sigma$-model, which 
takes account of some higher orders in the chiral expansion. 
Since the GP's $P^{\left(1 1, 1 1 \right)1}$ and 
$P^{\left(0 1, 1 2 \right)1}$ receive 
non-negligible contributions from the $\Delta$ and $D_{13}$
resonances \cite{Pas00}, a complete agreement with the ChPT results, 
dealing a priori only with non-resonant excitations, cannot be expected. 
On the other hand, the resonance contribution to 
$P^{\left(0 1, 0 1 \right)1}$ is extremely small,
and therefore the large deviation of the leading order ChPT result 
indicates the necessity of a higher order calculation.
\newline
\indent
In conclusion, we have presented a DR formalism for VCS and 
given, for the first time, a dispersive result for 4 of the 
GP's of the proton. These evaluations could be used to check the 
convergence of the chiral expansions in ChPT calculations.
It also provides a new tool to analyze VCS experiments at
higher energies (above pion threshold) where there is an increased
sensitivity to the GP's of the proton. 
In the future we will study further details of such an
analysis, including possible parametrizations of the 2 non-convergent
dispersion integrals. 
\newline
\indent
This work was supported by the ECT*, the Deutsche
Forschungsgemeinschaft (SFB443), 
and the EU/TMR Contract No. ERB FMRX-CT96-0008.

\end{document}